\documentclass[a4paper,twocolumn,11pt,accepted=2025-04-07]{quantumarticle}
\pdfoutput=1
\usepackage[table]{xcolor}
\usepackage{float}
\usepackage[utf8]{inputenc}
\usepackage[english]{babel}
\usepackage[T1]{fontenc}
\usepackage{amsmath}
\usepackage{hyperref}
\usepackage[numbers,sort&compress]{natbib}

\usepackage{tikz}
\usepackage{lipsum}
\usepackage{bbm, amsthm, bm,textcomp, nicefrac,geometry,ragged2e}
\usepackage{multicol}
\usepackage[bbgreekl]{mathbbol}
\usepackage{graphicx,epstopdf,color,verbatim,enumitem,ulem}
\usepackage{pbox,array}
\usepackage{mathrsfs}
\usepackage{amssymb}
\usepackage{mathtools}
\usepackage{stackrel}
\usepackage[thinlines]{easytable}
\usepackage{amsmath}
\makeatletter
\usepackage[caption=false]{subfig}
\usepackage[T1]{fontenc}
\usepackage[caption=false]{subfig}
\usepackage{babel}
\addto\captionsenglish{}

\usepackage{amsthm}
\usepackage{amsmath}
\usepackage{thmtools}

\usepackage[linesnumbered,ruled]{algorithm2e}
\usepackage{hyperref}

\hypersetup{
    colorlinks = true,
    linkcolor = teal,
    citecolor = teal,
    filecolor =  teal,      
    urlcolor =  teal,
}

\newcommand{\ket}[1]{\left|#1\right\rangle}

\newcommand{\bra}[1]{\left\langle #1\right|}

\newcommand{\proj}[1]{\ket{#1}\!\bra{#1}}

\newcommand{\px}{\sigma_x}

\newcommand{\pz}{\sigma_z}
\newcommand{\id}{\mathbb{1}}

\begin{document}

\sloppy

\title{Improving entanglement purification through coherent superposition of roles}

 \author{Jorge Miguel-Ramiro}
 \author{Alexander Pirker}
 \author{Wolfgang D\"ur}

\affiliation{Universit\"at Innsbruck, Institut f\"ur Theoretische Physik, Technikerstra{\ss}e 21a, 6020 Innsbruck, Austria}

\begin{abstract}
Entanglement purification and distillation protocols are essential for harnessing the full potential of quantum communication technologies. Multiple strategies have been proposed to approach and optimize such protocols, most however restricted to Clifford operations. In this paper, we introduce a superposed entanglement purification design strategy, leveraging coherent superpositions of the roles of entangled states to enhance purification efficiency, defining a new family of non-Clifford distillation protocols. We demonstrate how this approach can be hierarchically integrated with existing entanglement purification strategies, consistently improving protocols performance. 

\end{abstract}

\maketitle

\section{Introduction}
\label{sec:intro}

Quantum communication stands at the forefront of emergent quantum technologies, promising unique and powerful applications such as quantum cryptography \cite{Ekert91, Lo2014}, distributed quantum computing \cite{CiracDistributed, Hayashi15,Cacciapuoti2020}, distributed sensing \cite{Kessler2014, Eldredge2018,Sekatski2020}, or clock synchronization \cite{ClockSin}.  At the core of these applications lies entanglement, demanding the development of tools that enable faithful and high-quality generation, distribution, storage, and manipulation of such entanglement.

Among these tools, entanglement purification emerges as a fundamental technique to counteract the adverse effects of noise and decoherence, crucial for effectively distributing entanglement in applications such as quantum repeaters \cite{Briegel1998,Duer1999} quantum key distribution \cite{QKD}, and quantum secret sharing \cite{Hillery1999,Gottermansecret}. It provides an alternative to other approaches like quantum error correction \cite{knill97,Duer2007} in mitigating noise.

Entanglement purification protocols (EPPs) essentially involve distilling a few high-quality entangled states from a larger number of lower-quality copies, achieved by using only local operations at each involved party. Different strategies have been proposed, including recurrence \cite{Bennett1996,Deutsch_1996,Verstraete05,Fang2020} and hashing-like \cite{Bennett1996, Verstraete05, ferranPRL, ferranPRA} protocols, that work for bipartite \cite{Bennett1996,Deutsch_1996,Bennett19962,Bombin2006,Duer2007,Ruan2018,Bauml2018,Das2020,Hu2021} or multipartite \cite{M7,Smolin2005, multipartiterecurrence2dim,smolinmaneva} entanglement, as well as for qudits \cite{Horodecki_1999,Alber_2001,MiguelRamiro2018} or other kinds of entanglement \cite{Eisert_2004,Miyake2005,Hage_2008, Fiur2010, Miguel_W}. All of them have in common an inner working mechanism, that consists in transferring information from some copies into other one(s). These copies are subsequently measured to learn some information about the remaining states, eventually increasing their fidelity, either in a probabilistic or a deterministic way.

EPPs have undergone extensive investigation, yielding optimal constructions in some cases \cite{Fujii2009,Rozp2018,Krastanov_2019}. However, concrete protocol proposals are predominantly confined to Clifford operations and only a few efforts have explored potential advantages of protocols beyond Clifford circuits \cite{Torres2016,Preti2022,Yan_2023,Preti2024}, largely due to the complexity of optimizing and identifying suitable constructions within larger classes of operations.   

In this work, we propose an alternative strategy inspired by approaches from diverse quantum information contexts \cite{MiguelRamiro2018, Chiribella2019, Abbott2020, Kristjnsson_2020, RB_2021, Rubino2021}, exploring the unique advantages of channel superposition when applied to basic existing entanglement distillation strategies. Fundamentally, this approach enables diverse EPP configurations to operate simultaneously in coherent superposition. As a side effect of our approach, we naturally introduce a  new family of non-Clifford entanglement purification protocols. These protocols can simplify the challenge of searching for solutions in expansive operation sets, particularly those that lie outside the Clifford group.

As we demonstrate, superposition can be achieved using a quantum control plane, inherently provided by one or more additional copies of the noisy initial entangled state, ensuring the resource overhead remains practical. This control plane facilitates coherent state-role exchanges or complete protocol exchanges in superposition. While this increases resource demands, we show that the recursive nature of our approach consistently outperforms existing methods and protocols. Importantly, the observed performance enhancements are not only significant under ideal conditions but also robust in noisy operational regimes and for different kinds of initial states.

Our method provides a versatile framework for combining arbitrary existing EPPs to construct distillation protocols that can operate with enhanced performance. Moreover, this strategy effectively introduces a direct way to design alternative families of entanglement purification protocols, which are not limited to specific cases. In particular, one can apply our approach to any known entanglement purification protocol --thereby improving its performance--, but also to ones which will be developed in the future.

This work is structured as follows. In Sec.~\ref{sec:background}, we review the relevant notions and tools utilized in this paper. Sec.~\ref{sec:protocols} introduces our proposed role-exchanging superposed method, wherein we analyze and compare its performance, including noisy conditions. The extension of these ideas is discussed in Sec.~\ref{sec:more copies}. We summarize and conclude in Sec.~\ref{sec:conclusions}.

\section{Background}
\label{sec:background}
We briefly review here the basic concepts and notations we use throughout this work.

\subsection{Bell states}
Bell states, also known as EPR states, are two-qubit maximally entangled quantum states, mathematically defined as
\begin{align}
\ket{\mathrm{\Phi}_{ij}} = (\id \otimes \px^j \pz^i) \ket{\mathrm{\Phi}_{00}}, \label{eq:bellbasis}
\end{align}
where $i,j = ( 0,1 )$, and $\ket{{\Phi}_{00}} = (\ket{00} + \ket{11})/\sqrt{2}$ is the reference Bell state we take in this work. The operators $\{ \sigma_z, \sigma_x, \sigma_y  \}$ correspond to the Pauli matrices \cite{nielsen_chuang_2010}.

\subsection{Noisy channels and state fidelity}
The distribution, storage and manipulation of quantum entanglement is an in-practise complicated task due to the unavoidable detrimental effects of noise and decoherence, that jeopardize the quality of the entanglement. We represent the effect of such decoherence by completely positive (CP) quantum maps that deviate perfectly pure Bell states into probabilistic mixtures of states.

Any quantum CP noisy channel can be described using the Kraus operator representation \cite{nielsen_chuang_2010}, i.e., 
\begin{equation}
     \xi (\rho) = \sum_i K_i \rho K_i^{\dagger}, \label{eq:kraus}
\end{equation}
where $\sum_i K_i^{\dagger} K_i = \id$.

In particular, we consider different  CP noisy channels that affect the initially distributed states to be purified, and also model imperfect protocol operations. The choice of these channels is motivated by the actual modeling of current quantum technologies \cite{Georgopolus2021}. This includes very common errors, such as Pauli errors, which map the reference state  $ \ket{\mathrm{\Phi}_{00}}$  into other basis states. An example is the dephasing channel $\xi_{\rm deph}$,
 \begin{equation} \label{eq:deph}
\text{Dephasing:}\:\: \thinspace K_0 = \begin{pmatrix}
1 & 0 \\
0 & 1
\end{pmatrix}= \id, \:
K_1 = \begin{pmatrix}
1 &  0\\
0 & -1 
\end{pmatrix} = \sigma_z.
\end{equation}
Another particularly relevant noise model is the depolarizing noise $\xi_{\rm depo}$ \cite{nielsen_chuang_2010} (or white noise), 
\begin{align}\label{eq:depo}
     \text{Depolarizing:}\:\: K_0=\id, K_1=\sigma_z, K_2=\sigma_x,K_3=\sigma_y.
\end{align}
When depolarizing noise affects one or two parties of the Bell state $\ket{\mathrm{\Phi}_{00}}$, the resulting state is called Werner state \cite{nielsen_chuang_2010}, 
\begin{equation}
\rho= F \proj{\Phi_{00}} + \frac{1-F}{3} \sum_{i}( \sigma_i \otimes \id) \proj{\Phi_{00}}  (\sigma_i \otimes \id),
 \label{eq:depolarizing_intro}
\end{equation} 
where $\sigma_i =\{\sigma_{z},\sigma_{x},\sigma_{ y} \}$ for $i=\{1,2,3\}$. This equals 
\begin{equation}
 \rho= q \proj{\Phi_{00}} + \frac{1-q}{4} \id, 
 \label{eq:depolarizing_intro2}
\end{equation}
with $F=\frac{3q+1}{4}$. This represent a worst-case scenario (white noise), since any other state can be brought to this Werner form with local operation and without altering its fidelity (although partially destroying some entanglement), by means of depolarization techniques \cite{Dur2005}. Finally, we also consider amplitude damping (or decay) channel $\xi_{\rm damp}$, given by  
 \begin{equation}\label{eq:damp}
\text{Damping:} \thinspace \thinspace K_0 = \begin{pmatrix}
1 & 0 \\
0 & \sqrt{1 - \gamma}
\end{pmatrix}, \thinspace\thinspace
K_1 = \begin{pmatrix}
0 & \sqrt{\gamma} \\
0 & 0
\end{pmatrix}.
\end{equation}
where $\gamma$ is the damping strength probability.

The fidelity of two quantum states measures the distance between those states. Since our reference state is always the $\ket{\Phi_{00}}$, it suffices to define the fidelity of an arbitrary mixed state $\rho$ with respect to $\ket{\Phi_{00}}$ as $F=\bra{\Phi_{00}} \rho \ket{\Phi_{00}}$. Note that the fidelity of a Werner state exactly equals $F$ in Eq.~(\ref{eq:depolarizing_intro}).

\subsection{Entanglement purification protocols (EPPs)}  
Entanglement purification and distillation protocols \cite{Duer2007} encompass a set of strategies aimed at taking a set of noisy entangled copies shared between two or more parties and, through only local operations at each site, outputting a smaller number of copies with enhanced fidelity.

\begin{figure} 
    \centering
    \subfloat[\centering]
  {\includegraphics[width=0.5\columnwidth]{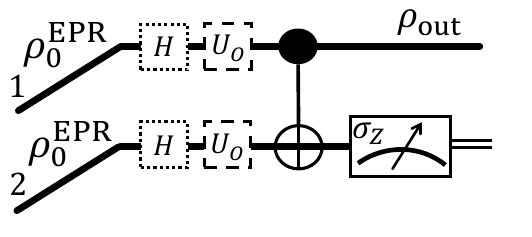} \label{fig:singleSEL}}
     \subfloat[\centering]{\includegraphics[width=0.5\columnwidth]{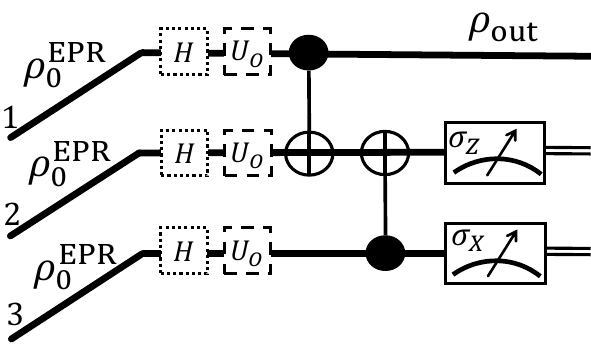} \label{fig:doublesel}} 
    \caption{\label{fig:singleanddouble} (a) Single selection protocol equivalent to the BBPSSW EPP \cite{Bennett1996}, that is equivalent to the Oxford EPP \cite{Deutsch_1996} when introducing $U_O$ rotation. Alice and Bob perform the same operations and output is kept if coincident outcomes are found in the measurements. A local Hadamard rotation $H$ can be adaptability included (P1-P2 variation) to further enhance the protocols performance. (b) Double selection protocol initially proposed in \cite{Fujii2009}, including the Oxford and P1-P2 variations. }
\end{figure}

Given the conceptual scope of this work, we primarily focus on recurrence EPPs, initially introduced in \cite{Bennett1996, Deutsch_1996}. In its basic setting, the BBPSSW approach \cite{Bennett1996} (see Fig. \ref{fig:singleSEL}) iteratively operates on two noisy entangled states. Basic two-qubit operations are performed at each site between the copies, effectively transferring some information from one copy to the other. Subsequent measurement of the latter copy enables one to effectively learn some information about the former, resulting in a probabilistic increase in the fidelity of the surviving state. The selection process requires two-way classical communication. The Oxford (or DEJMPS) variant of the protocol \cite{Deutsch_1996} includes a local unitary rotation $U_O$ given by 
\begin{align}
&\ket{0}_{\rm A} \rightarrow \tfrac{1}{\sqrt{2}} (\ket{0}_{\rm A} - i\ket{1}_{\rm A}),  \ket{1}_{\rm A} \rightarrow \tfrac{1}{\sqrt{2}} (\ket{1}_{\rm A} - i\ket{0}_{\rm A}), \\
&\ket{0}_{\rm B} \rightarrow \tfrac{1}{\sqrt{2}} (\ket{0}_{\rm B} + i\ket{1}_{\rm B}),  \ket{1}_{\rm B} \rightarrow \tfrac{1}{\sqrt{2}} (\ket{1}_{\rm B} + i\ket{0}_{\rm B}).
\end{align}

In addition, the so-called P1-P2 variation of both EPPs includes an adaptive application of the local Hadamard rotation, $
H = \frac{1}{\sqrt{2}}
\begin{pmatrix}
1 & 1 \\
1 & -1
\end{pmatrix}
$. This enables the definition of two subroutines, P1 and P2, which operate with and without the Hadamard gate, respectively, each targeting the detection of different types of errors. These subroutines can be iteratively alternated or optimally selected to further enhance the performance of the protocols.

Further improvements to this basic setting have been proposed \cite{Fujii2009,Rozp2018,Krastanov_2019,Preti2022,Yan_2023,Preti2024}, including those involving more copies. The double selection protocol \cite{Fujii2009}, depicted in Fig. \ref{fig:doublesel}, stands out as one of the best extensions utilizing only three raw copies.

The performance of EPPs can be evaluated by considering various figures of merit, each with individual relevance for different practical considerations. These evaluation parameters can include fidelity convergence speed, protocol success probability, or general efficiency (yield), defined as 
\begin{equation}
     {\rm Y}=\frac{\# \, \rm copies \, \rm obtained} {\# \, \rm copies \, \rm consumed}, \label{eq:yield}
\end{equation}
that accounts for the resource overheads required to complete the protocol, i.e., it evaluates the efficiency of the protocol. Rigorously, the value of the amount of copies obtained is restricted to an
accuracy condition, so that the protocol is considered successful if the final fidelity satisfies $F > 1-\epsilon$. For recurrence protocols with a single copy as output, the obtained number of copies is given by $\frac{\prod_k P_k}{m^k}$, where $P_k$ is the success probability of the $k^{\rm th}$ iteration and $m$ is the number of copies used in each round.

\subsection{Superposition of quantum channels}
The concepts behind this work are inspired by recent works demonstrating the advantages of implementing quantum channels in coherent superposition, also known as superposition of trajectories. The distinctive correlation features emerging from these superposition processes have shown benefits in various domains of quantum computing \cite{RB_2021, SQEM1, SQEM2} and quantum communication \cite{MiguelRamiro2021, Chiribella2019, Abbott2020, Kristjnsson_2020, Rubino2021}. This body of research underscores the potential of leveraging coherence in different quantum protocols to achieve enhanced performance, applicability and efficiency. 

Superposition of different tasks is achieved by introducing a quantum control register in a superposition state, which acts as a control to exchange roles of states or tasks via a controlled-SWAP operation, also known as Fredkin gate \cite{Fredkin82,Patel_2016}, which is a non-Clifford operation described as
\begin{equation}\label{eq:GB-QCcswap}
\begin{split}
      \text{cSWAP} = & \ket{0} \bra{0}_{\rm c} \otimes  \id  + \sum^{d-1}_{i=1}  \ket{i} \bra{i}_{\rm c} \otimes \text{SWAP}_{{\rm a}, {\rm b}_{i}},
  \end{split}
\end{equation}
for a $d$-dimensional control register, such that when applied, e.g., to $\ket{+}_{\rm c} \ket{\phi}_{\rm A} \ket{\psi}_{\rm B}$, the registers $\rm A,B$ are coherently swapped depending on the state of the control qubit, i.e. 
\begin{align}\label{eq:GB-QCcswap2}
      &\text{cSWAP} \ket{+}_{\rm c} \ket{\phi}_{\rm A} \ket{\psi}_{\rm B} = \\ \notag
      &\frac{1}{\sqrt{2}} \left( \ket{0}_{\rm c} \ket{\phi}_{\rm A} \ket{\psi}_{\rm B} +\ket{1}_{\rm c} \ket{\psi}_{\rm A} \ket{\phi}_{\rm B} \right).
\end{align}

\section{Superposed EPP: role exchange}
\label{sec:protocols}
We present and examine various protocols that utilize the concept of superposing entanglement purification subroutines. These protocols serve as proof-of-principle strategies, highlighting the advantages of such implementations. However, further refinement is necessary to achieve optimal results.

\subsection{Basic protocol} \label{sec:Role exchanging}
We initially propose employing a strategy utilizing the basic recurrence entanglement purification and distillation protocol, i.e., BBPSSW \cite{Bennett1996, Bennett19962}, operating on two identical mixed states. By introducing an additional copy, taking the role of the control state, we establish an effective $3 \rightarrow 1$  purification protocol. As demonstrated later, the protocol yields advantages even when the control state exhibits identical or even increased noise levels compared to the other states of the ensemble. This highlights the robustness and efficacy of the protocol in diverse scenarios.

Our strategy revolves around utilizing three noisy copies as inputs for the protocol. One of these copies, designated as the control copy, dictates the roles of the remaining two states, resulting in a coherent exchange where each acts as the target of the purification rounds based on the state of the control copy, via a superposition of the different role configurations. Upon completing the protocol, the control copy is measured in the $\sigma_x$ basis, and the favorable outcomes are post-selected, leading to enhanced protocol performance.

For clarity in the analytical analysis, we assume a perfect Bell state as the control state. This assumption is relaxed in the performance numerical analyses.

The protocol encompasses the following steps (see also Fig. \ref{fig:protocol1}):

\begin{figure}[ht]
    \centering
    \includegraphics[width=0.93\columnwidth]{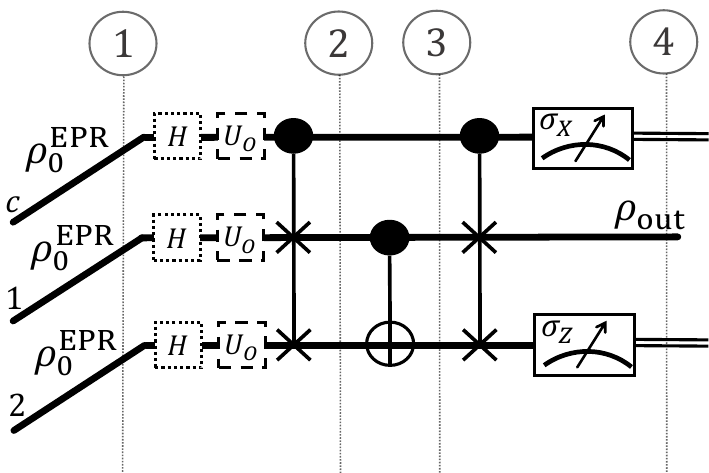}
    \caption{\label{fig:protocol1} Circuit illustration of the role-exchanging superposed EPP, possibly including the Oxford ($U_O$) and P1-P2 ($H$) variations. Only Alice's side is shown, Bob's operations are identical.  }
\end{figure}

   \begin{enumerate}

 \item Three copies of entangled states are considered. One of them plays the role of superposition-control state ('$\rm c$'), and we model the initial noise of the other two copies as white noise (depolarizing). We remark that the control copy is assumed to be perfect for the analytical analysis but can be replaced for a noisy one identical to the other two (see below). We represent the states in the Pauli basis, such that
     \begin{equation}
   \begin{aligned}
     & \ket{\Phi_{00}}\bra{\Phi_{00}}_{\rm A_{\rm c} B_{\rm c}} \\ 
     & \otimes  \left(\sum_{i,j=0}^1  {p}_{i,j}  \Sigma_{i,j}  \ket{\Phi_{00}} \bra{\Phi_{00}}_{\rm A_1 B_1} \Sigma^{\dagger}_{i,j} \right) \\ 
       &\otimes   \left( \sum_{q,r=0}^1  {p}_{q,r} \Sigma_{q,r}  \ket{\Phi_{00}}\bra{\Phi_{00}}_{ \rm A_2 B_2} \Sigma^{\dagger}_{q,r}\right),
   \end{aligned}
       \end{equation}

where we can assume without loss of generality that $\Sigma_{ij} \equiv \id_{\rm A}\otimes {\sigma_{ij}}_{\rm B}$, where $\{\sigma_{00},\sigma_{01},\sigma_{10},\sigma_{11} \}=\{\id,\sigma_{z},\sigma_{x},\sigma_{ y} \}$, such that  $\Sigma_{ij} \ket{\Phi_{00}} =  \ket{\Phi_{ij}}$ maps all the basis Bell states. Here $ {p}_{i,j}$ determines the strength of each noisy contribution, where $ {p}_{00}$ corresponds to the state initial fidelity.


\item The two copies ('$1$' and '$2$') undergo a coherent swapping depending on the state of the control. This is achieved by local controlled-swap operations (cSWAP), also know as quantum Fredkin operation \cite{Fredkin82,Patel_2016}, ${\rm cSWAP_{c \rightarrow 12}}$, performed at each site/party, that generate a superposition between the states, such that each state will play a different purification role (target state) in each branch of the superposition. The global state reads
           \begin{widetext}
    \begin{equation}
    \begin{aligned}
        \quad\quad\quad\quad\quad\quad &\ket{{ss}}\bra{{ss}}_{\rm A_{\rm c} B_{\rm c}} \otimes \left(\sum_{i,j,q,r=0}^1  { p}_{i,j}  \ket{\Phi_{i,j}} \bra{\Phi_{i,j}}_{\rm A_1 B_1} \otimes {p}_{q,r}  \ket{\Phi_{q,r}}\bra{\Phi_{q,r}}_{\rm A_2 B_2} \right) \\
         +& \frac{1}{2} \ket{{00}}\bra{{11}}_{\rm A_{\rm c} B_{\rm c}} \otimes \left(\sum_{i,j,q,r=0}^1  {p}_{i,j}  \ket{\Phi_{i,j}} \bra{\Phi_{q,r}}_{\rm A_1 B_1} \otimes {p}_{q,r} \ket{\Phi_{q,r}} \bra{\Phi_{i,j}}_{\rm A_2 B_2} \right) \\
         +& \frac{1}{2} \ket{{11}}\bra{{00}}_{\rm A_{\rm c} B_{\rm c}} \otimes \left(\sum_{i,j,q,r=0}^1  {p}_{i,j}  \ket{\Phi_{q,r}} \bra{\Phi_{i,j}}_{\rm A_1 B_1} \otimes {p}_{q,r} \ket{\Phi_{i,j}} \bra{\Phi_{q,r}}_{\rm A_2 B_2} \right),
    \end{aligned}
    \end{equation}
    \end{widetext}

      where $ \ket{{ss}}\bra{{ss}}$ refers to both the $ \ket{{00}}\bra{{00}}$ and $ \ket{{11}}\bra{{11}}$ terms.
    \item Next, following the BBPSSW protocol, bilateral cNOT operations (bcNOT) are applied locally by parties $\rm A$ and  $\rm B$  from copy $1$ to copy $2$, leading to

 \begin{widetext}
 \begin{equation}
  \begin{aligned}
       \quad\quad\quad\quad &\ket{{ss}}\bra{{ss}}_{\rm A_{\rm c} B_{\rm c}} \otimes  \left(\sum_{i,j,q,r=0}^1  { p}_{i,j}  \ket{\Phi_{i \oplus q,j}} \bra{\Phi_{i \oplus q,j}}_{\rm A_1 B_1} \otimes    {p}_{q,r}  \ket{\Phi_{q,r\oplus j}}\bra{\Phi_{q,r\oplus j}}_{ \rm A_2 B_2} \right)  \\ 
       +&\frac{1}{2}\ket{{00}}\bra{{11}}_{\rm A_{\rm c} B_{\rm c}} \otimes  \left(\sum_{i,j,q,r=0}^1  {p}_{i,j}   \ket{\Phi_{i\oplus q,j}} \bra{\Phi_{q\oplus i,r}}_{\rm A_1 B_1}  \otimes     {p}_{q,r} 
       \ket{\Phi_{q,r\oplus j}}\bra{\Phi_{i,j\oplus r}}_{ \rm A_2 B_2} \right)  \\ 
       +& \frac{1}{2}\ket{{11}}\bra{{00}}_{\rm A_{\rm c} B_{\rm c}} \otimes  \left(\sum_{i,j,q,r=0}^1  {p}_{i,j}    \ket{\Phi_{q\oplus i,r}} \bra{\Phi_{i\oplus q,j}}_{\rm A_1 B_1}  \otimes    {p}_{q,r} \ket{\Phi_{i,j\oplus r}}\bra{\Phi_{q,r\oplus j}}_{ \rm A_2 B_2} \right).
   \end{aligned}
   \end{equation}
    \end{widetext}

    \item Subsequently, another controlled-SWAP (cSWAP) operation can be implemented, although this step is \textit{not} necessary. Finally, the second copy is measured locally in the $Z$ basis at each site. If parties $A$ and $B$ obtain the same outcome ($j \oplus r=0$), the first state is retained.  The only components that survive the measurement are then the ones coming from $\left(  {p}_{q,j} \ket{\Phi_{i,0}}\bra{\Phi_{q,0}}_{ \rm A_2 B_2} \right)$. Simultaneously, the control state is measured locally in the $X$ basis by $\rm A$ and  $\rm B$, post-selecting the desired outcomes as stated in the following.


   %
   \end{enumerate}

If the measurement outcomes of $A$ and $B$ of the second copy match and correspond to $0$, then regardless of the control register's outcomes, the fidelity of the surviving state equals the fidelity found in the BBPSSW protocol, where all branches lead to the same result. However, if the outcome found is $(1,1)$ in the second copy and  $\{(+,+), (-,-)\}$ in the control register, factors corresponding to $\{(q,i)\}=\{(0,1), (1,0)\}$ acquire a "minus" phase in the coherent terms, resulting in vanishing terms and further enhancement of the component $\ket{\Phi_{00}}$ (i.e., of the fidelity). The resulting state in this case reads
 \begin{equation}
  \begin{aligned}
       \frac{1}{P} \left( \sum_i { p}^{2}_{i 0}  \ket{\Phi_{00}} \bra{\Phi_{00}}_{\rm A_1 B_1} +\sum_i { p}^{2}_{i 1} 
 \ket{\Phi_{01}} \bra{\Phi_{01}}_{\rm A_1 B_1} \right),
   \end{aligned}
   \end{equation}
where $ P = \sum_j { p}^{2}_{jj}$ is a normalization constant that equals the success probability, and where the new fidelity is $F=\dfrac{\sum_i { p}^{2}_{i 0}}{\sum_j { p}^{2}_{jj}}$.

Notice that the only error left uncorrected in the previous step corresponds to states from the terms $\ket{\Phi_{01}}$. With this observation in mind, one can advance to the next iteration of the protocol after performing a $(H \otimes H)$ operation with $H$ the Hadamard operator, that is, \textit{without} the need to depolarize again into a Werner-like state. This operation interchanges components associated with $\ket{\Phi_{01}}$ and $\ket{\Phi_{10}}$, given that the latter error is rectified by the protocol. 

Two protocol iterations then ensure the attainment of a perfectly purified copy. However, it is crucial to remark that in this process, we operate under the assumption of a perfect Bell state controlling the superposition, which results in an overall negative protocol efficiency. We remark that the superposed role-exchanging of the protocol it is not restricted to initial Werner states and does not require depolarization after each round.

{\LinesNumberedHidden
    \begin{algorithm}
        \SetKwInOut{Input}{Input}
        \SetKwInOut{Output}{Output}
        \SetAlgorithmName{Algorithm}{}
 \justifying \textit{Input}: Three copies of a noisy bipartite entangled state $\rho_0$. Alice and Bob perform identical operations.
        \begin{enumerate}
            \item Prepare three copies of a noisy entangled state $\rho_0$.
            \item Apply suitable bilateral rotations $(H, U_0)$ depending on the protocol variant (Oxford, P1, P2). This can be optimized such that the best variant is chosen in each iteration.  Then, bilaterally apply a cSWAP operation with the first copy acting as superposition-control. 
            \item Apply a bilateral cNOT operation between the second and third copy.
            \item After a second cSWAP for recombination, measure the third copy in the Z basis and the control copy in the X basis. Keep the second copy if outcomes coincide for both measurements. 
        \end{enumerate}
\justifying  \textit{Output}: A purified entangled state with enhanced fidelity. A basic EPP protocol is executed but with the control and target roles coherently exchanged during the process.
\caption{Superposed EPP. Role exchange}
\end{algorithm}}

\begin{figure} 
    \centering
     \subfloat[\centering]{\includegraphics[width=0.9\columnwidth]{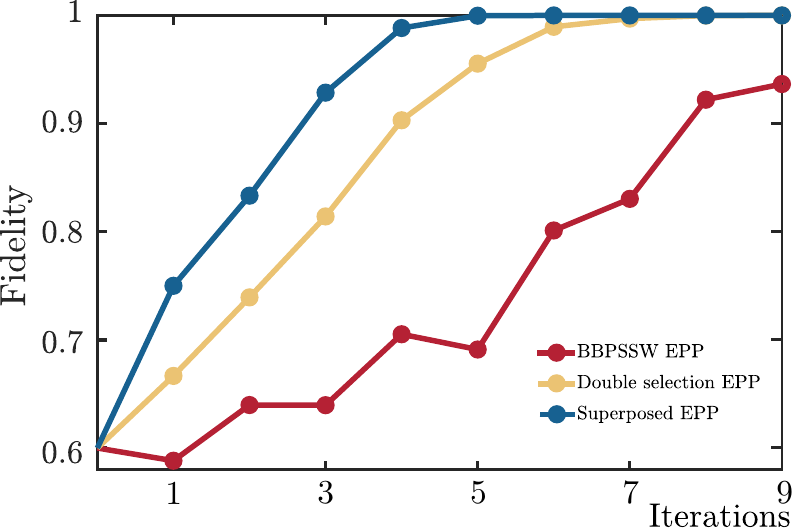} \label{fig:role_c}} \\
     \subfloat[\centering]{\includegraphics[width=0.9\columnwidth]{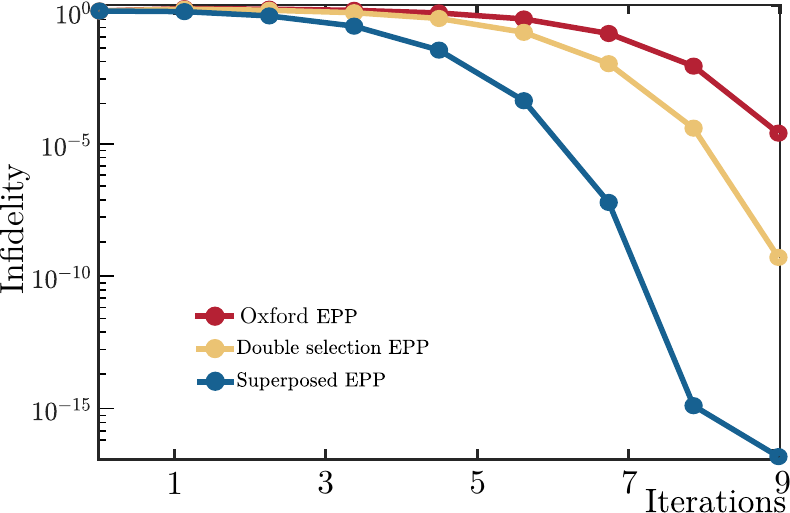} \label{fig:role_a}} 
    \caption{\label{fig:role1} (a) Fidelity evolution as a function of the number iterations for the BBPSSW P1-or-P2 EPPs with initial states affected (both sides) by amplitude damping noise of strength $q=0.6$.  (b) Infidelity evolution as a function of the number iterations for the different P1-or-P2 Oxford variation protocols with initial states affected by dephasing noise and fidelity $F_0=0.6$  (see Appendix \ref{sec:appendixA} for a detailed analysis of each protocol for correcting Pauli noises). In all cases, depolarizing the initial states into Werner form decreases the overall protocols performances. }
\end{figure}

To address this, we turn our attention to a realistic scenario: a $3 \rightarrow 1$ protocol involving a control copy identically noisy as the others, as analyzed in Fig.~\ref{fig:role1}. We show how the role-exchanging superposed entanglement purification approach introduced above, can significantly surpass single selection protocols \cite{Bennett1996, Bennett19962, Deutsch_1996}, and also more elaborated ones such as the double selection protocol, Fig.~\ref{fig:singleanddouble},  \cite{Fujii2009}, which is considered among the most efficient EPPs modifications using $3$ raw EPR copies \cite{Krastanov_2019}. We consider different initial entangled states, affected by amplitude damping in Fig.~\ref{fig:role1} (a), and dephasing noise in Fig.~\ref{fig:role1} (b).  In addition, we consider the P1-or-P2 variants of all the protocols (see Figs.~\ref{fig:singleanddouble}~-~\ref{fig:protocol1}), meaning that the best subroutine (P1 or P2) is selected in each iteration. We analyze the performance for different kinds of initial states. The plots show the increase in fidelity $F$ (or decrease in infidelity, defined as $1-F$) with the number of steps, where one observes that the superposed protocol performs better than the elementary ones.

We provide in  Appendix \ref{sec:appendixA} a detailed analysis of the density matrix evolution of the superposed EPP protocol compared with the Oxford and double selection strategies, for different instances of initial states, including single Pauli noise and Werner states. 

Although more involved protocols have been investigated \cite{Rozp2018, Krastanov_2019}, usually relying on more raw copies and intended for optimizing certain figures of merit, our results underscore the potential of our strategy, which can be conceived as an efficiency booster for any existing strategy. We elaborate below on how this concept can be generalized by leveraging efficient existing protocols as building blocks and applying similar superposition techniques on top of them.

\subsection{Imperfect operations}
Any realistic implementation of an entanglement purification and distillation protocol would inevitably be subjected to noise stemming from the imperfect execution of the various operations involved. In particular, our superposed approach necessitates a certain overhead in resources that must be carefully considered, especially concerning the controlled-swap (cSWAP) gate.
\begin{figure} 
    \centering
    \subfloat
  {\centering\includegraphics[width=0.47\textwidth]{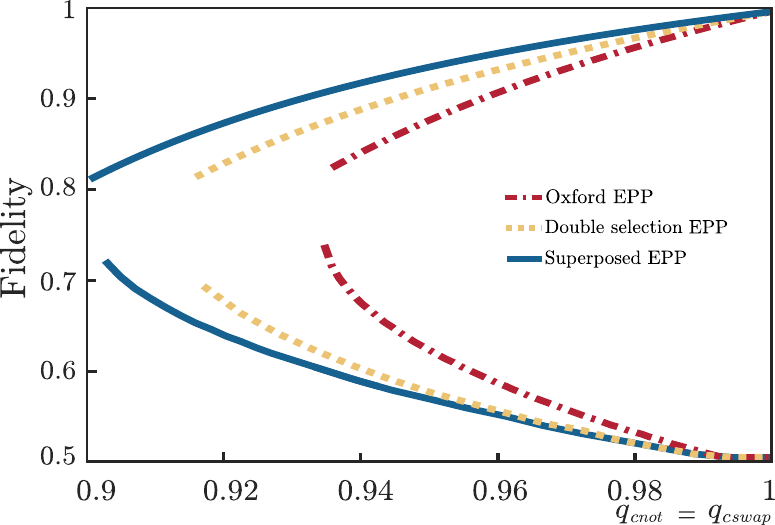} \label{fig:noisy1}} \vfill
    \subfloat
  {\centering\includegraphics[width=0.47\textwidth]{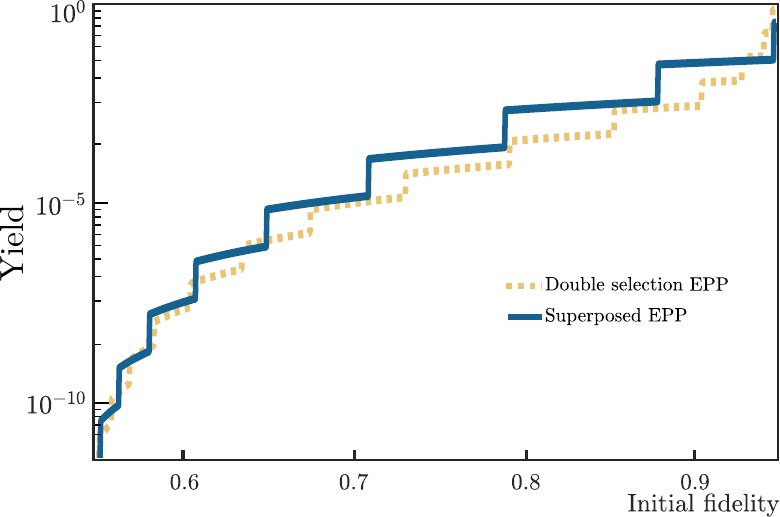} \label{fig:noisy2}}
    \caption{\label{fig:noisy} (a) Operational regime of the different protocols under imperfect operations, where minimum required and maximum achievable fidelities are shown as a function of the noise parameter. Operational noise is modeled by local amplitude damping noise with parameter $q$ acting on each qubit involved in a gate after each gate is implemented. (b) Yield for the Oxford variation with $3\%$ of operational damping noise, i.e., $q_{\rm cnot}=q_{\rm cSWAP}=0.97$ and initial Werner states with $F_{\rm target} =0.95$. The single selection protocol is not shown since never reaches $F_{\rm target}$.}
\end{figure}

The cSWAP operation, a non-Clifford gate also known as the Fredkin gate \cite{Fredkin82, Patel_2016}, has been extensively explored in the literature, and various experimental proposals for its high-fidelity implementation already exist \cite{FredEX1A, FredEx1, FredEx1b, FredEx2, FredEx2b, FredEx3, FredEx3b}. These proposals are based on alternative strategies that circumvent the limitations of decomposing the cSWAP gate into elementary gates. Alternatively, the superposed EPP scheme could be implemented in an interferometric-based way \cite{Rubino2021, SQEM2}, circumventing those problems and possibly enhancing the quality of the implementation of such a gate.

Here, we analyze  the operational performance of the superposed role exchanging EPPs proposed before under the influence of noisy operations. We model the noise \cite{Duer2007} of the quantum gates involved as the ideal implementation of the unitary operation $U$, followed by some noisy channel acting on each of the involved qubits. That is, for a $k$-qubits imperfect unitary $\mathcal{E}_U $,
\begin{equation}
    \mathcal{E}_U (\rho) =    \xi^{(1)} \dots \xi^{(k)} \left( U \rho U^{\dagger}\right) ,
\end{equation}
where different choices of $\xi$ are considered. These choices include local depolarizing noise $\xi_{\rm depo}$ (affecting the qubits involved in each gate),  local amplitude damping noise $\xi_{\rm damp}$, Eq. (\ref{eq:damp}),  in Figs.~\ref{fig:noisy1}~-~\ref{fig:noisy2} and a combination of this with local dephasing noise  Eq.~(\ref{eq:deph}), following the modeling of current technologies gate noises \cite{Georgopolus2021}, covering a wide range of scenarios.  Measurement noise is modeled by POVM elements such that an orthogonal output is found with probability $p$, which for simplicity, is always chosen to be equal to the gate noise parameter $q$. 

Figs.~\ref{fig:noisy1}-~\ref{fig:noisy2} show the operational regime and yield of the protocols under operational amplitude damping noise, where the noise strength is assumed to be the same for the cNOTs ($q_{cnot}$) and cSWAPs operations ($q_{cSWAP}$). Notably, the superposed strategy, despite the noise from the cSWAP operations, tolerates higher levels of operational noise compared to the double selection protocol, both clearly surpassing more basic two-copy approaches. Additionally, the superposed EPP outperforms the double selection protocol in terms of fidelity evolution and overall performance.

\begin{figure}[ht]
    \centering
    \subfloat[\centering]{\includegraphics[width=0.9\columnwidth]{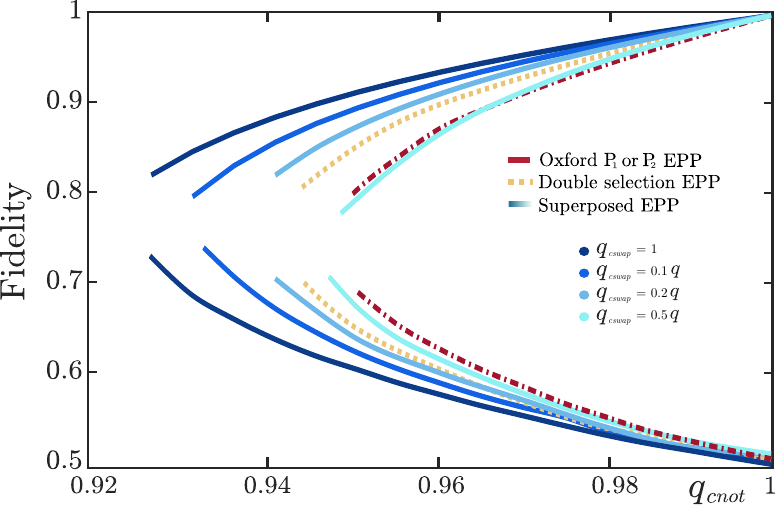} \label{fig:noisy3}} \\
    \subfloat[\centering]{\includegraphics[width=0.9\columnwidth]{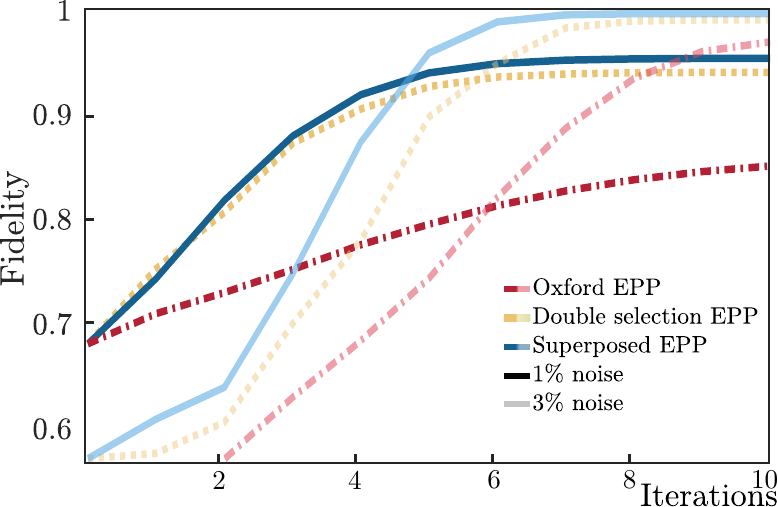} 
    \label{fig:noisy4}}
     \caption{\label{fig:noisy22}  (a) Operational regimes for Werner states and local depolarizing gate and measurement noise for different strengths of the cSWAP gate noise. (b) Fidelity evolution under noisy operations modeled by \cite{Georgopolus2021} ${\xi}_{\rm depo} \circ {\xi}_{\rm damp} \circ {\xi}_{\rm deph}$ with strength of $1\%$ (solid lines) and $3\%$ (shaded lines) for each noise.}
\end{figure}

Fig.~\ref{fig:noisy3} shows the operational regimes under operational local depolarizing noise and initial Werner states, for different values of the cSWAP noisy strength. We remark that the unique features of this gate could lead to alternative improved practical implementations \cite{FredEX1A, FredEx1, FredEx1b, FredEx2, FredEx2b, FredEx3, FredEx3b, Rubino2021, SQEM2}. 

Finally, we analyze in Fig.~\ref{fig:noisy4} the fidelity evolution for the different protocols under special operational noise considerations, motivated by realistic modeling of current quantum hardware \cite{Georgopolus2021}. Following such a model,  operational noise is defined as a concatenation of local depolarizing, damping, and dephasing noise affecting each qubit after each gate, i.e., $\xi_{\rm damp} \circ \xi_{\rm damp} \circ \xi_{\rm deph}$. The same noise parameters are used for every gate, consistently showing an advantage of the superposed EPP.

Relevant and consistent performance improvements of the superposed EPP are found under different considerations of initial states and noisy operations. 

%

\section{Further extensions} \label{sec:more copies}
The tools presented in this work can offer a recursive method for constructing more efficient EPPs, which can be seamlessly  integrated with existing entanglement purification strategies, by just implementing them in coherent superposition. This approach allows for the design of families of non-Clifford EPPs, where the controlled-swap operations serve as the non-Clifford extension of known protocols.

We illustrate and analyze various examples herein, emphasizing the advantage derived from our approach. Notably, this enhancement persists even when operating on more than three copies, a regime where previous studies \cite{Krastanov_2019} have shown only diminishing improvements in performance. Our results demonstrate that the superposed protocol continues to provide significant benefits beyond this point, offering a promising avenue for further optimization.

\begin{figure} 
    \centering
    \subfloat[\centering]
  {\includegraphics[width=0.45\columnwidth]{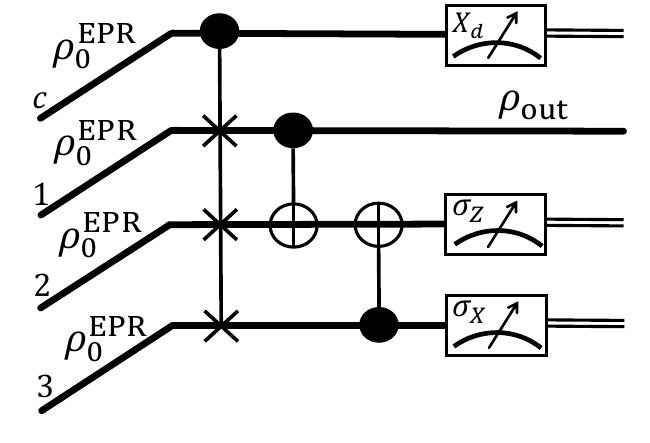} \label{fig:SUPdoublesel}}
     \subfloat[\centering]{\includegraphics[width=0.45\columnwidth]{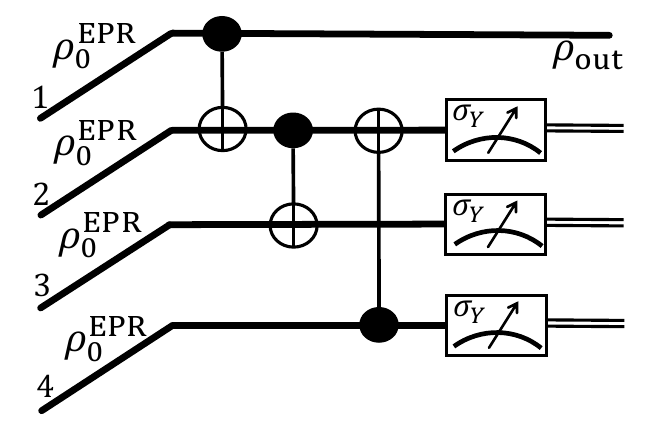} \label{fig:triplesel}} 
    \caption{\label{fig:singleandtriple} (a) Coherent superposed extension of the double selection EPP of Fig. \ref{fig:doublesel}. (b) Triple selection EPP from \cite{Krastanov_2019}. }
\end{figure}

Consider again the double selection protocol \cite{Fujii2009} illustrated in Fig.~\ref{fig:doublesel}. Following similar ideas to those used for the coherent extension for the two-copy protocol discussed in Sec.~\ref{sec:Role exchanging}, one can design a coherent version of the double selection protocol \cite{Fujii2009}. This can be done by integrating a control state that facilitates the construction of an effective $4 \rightarrow 1$ protocol, see Fig.~\ref{fig:SUPdoublesel}. It is important to note that the dimension of the control copy is now greater than two, necessitating careful consideration when assessing resources and protocol efficiency. 

The existing double selection strategy, Fig.~\ref{fig:doublesel},  involves three identical copies, of which two are eventually measured in different bases. Consequently, there are two target copies and three different roles during the protocol. For the superposed extension of such protocol, we first consider a three-dimensional control qubit (qutrit) state enabling us to coherently permute the surviving copy role. Subsequently, we explore a six-dimensional control copy allowing coherent permutations among all possible role configurations. Notice that these choices are made for illustrative purposes. In general, one can also construct higher-dimensional control entangled copies using the same ensemble qubit EPR states, effectively building control states of dimension $2^k$, where $k$ represents the number of entangled copies involved.

\begin{figure} 
    \centering
  {\includegraphics[width=0.95\columnwidth]{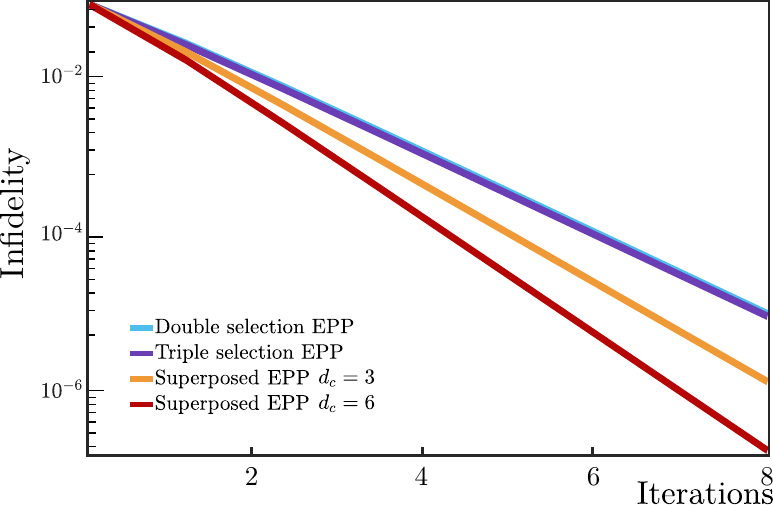} \label{fig:4copies_a}}
    \caption{\label{fig:4copies} Infidelity evolution as a function of the number iterations for the different protocols and initial Werner states with fidelity  $F_0=0.92$.  }
\end{figure}

Fig.~\ref{fig:4copies} shows the performance of the aforementioned superposed double selection protocol, utilizing a higher dimensional control copy initially prepared as a Werner state with fidelity $F_{\rm c}=F^{log_2(D)}$, where $D$ is the qudit dimension, incorporating this dimensionality factor for a fair resource comparison. The controlled-SWAP (cSWAP) operation facilitates the coherent swapping of roles among the three states designating the surviving one in each branch. Following this, the double selection protocol continues as usual, with the control copy measured now in the generalized Fourier basis for $d=D$, and the desired outcomes post-selected.

For comparison, we now include the triple selection protocol \cite{Krastanov_2019} (illustrated in Fig.~\ref{fig:triplesel}), a $4 \rightarrow 1$ protocol (triple selection) that only yields marginal performance improvement compared to the double selection approach, see \cite{Krastanov_2019} and Fig. \ref{fig:4copies}. Our superposed extension however demonstrates a clear ability to surpass this marginality, leading to further improvements that get even more relevant with higher dimension of the control register (i.e. larger number of superposed role exchanges).
We note that in Fig. \ref{fig:4copies} analysis we apply depolarization to Werner-like states after each round of the protocol. However, inspired by the analyses of the protocol discussed in Sec. \ref{sec:Role exchanging}, one could expect further performance improvements by allowing for analogous variants and modifications as the ones discussed in Sec. \ref{sec:Role exchanging}. 

\section{Conclusions}\label{sec:conclusions}

In this work, we have introduced an alternative strategy for entanglement purification and distillation based on the coherent superposition of protocols. By leveraging coherent permutations of the roles of entangled states in existing protocols, we have shown how this approach consistently enhances other protocols performance. Through comprehensive theoretical analyses and numerical simulations for different initial states and under imperfect operations, we have demonstrated the robustness and potential of our approach. Our findings indicate that the superposed strategy achieves higher fidelities when using more copies, also in the presence of operational noise, surpassing traditional strategies such as the double and triple selection methods.

This approach also enables the design of alternative entanglement purification protocols that effectively utilize non-Clifford operations, expanding the landscape of entanglement distillation. Furthermore, it offers a hierarchical framework that can be integrated with existing purification strategies, with consistent performance enhancements.

Preliminary results suggest that the tools introduced in this work could benefit other variants of EPPs as well. For instance, similar concepts could be applied to hashing techniques, where coherence tools can improve convergence into the set of likely sequences, thereby boosting protocol efficiency. We hope this work can open new avenues in the designing of enhanced strategies for advancing in the field of entanglement purification.

\section*{Acknowledgments}
This research was funded in whole or in part by the Austrian Science Fund (FWF) 10.55776/P36009 and 10.55776/P36010. For open access purposes, the author has applied a CC BY public copyright license to any author-accepted manuscript version arising from this submission.
\clearpage

\bibliographystyle{apsrev4-2}
\bibliography{superposedEPP.bib}

\clearpage
\onecolumngrid

\section*{Appendix A. Density matrix evolution for different EPP protocols} \label{sec:appendixA}
We show here the performance of different entanglement purification protocol as a function of the evolution of each of the Pauli diagonal elements during protocol iterations. In all protocols, the states remain Pauli diagonal during their execution. We consider the Oxford protocol (Fig. \ref{fig:singleanddouble} (a)), the double selection EPP (Fig. \ref{fig:singleanddouble} (b)),  and our superposed EPP (Fig. \ref{fig:protocol1}). Initial states considered vary for each row, including only Pauli $\sigma_x, \sigma_y, \sigma_z$ noise for the first three rows  respectively, and initial Werner states in the last one.

One can observe how the $3 \rightarrow 1$ protocols equal or outperform (i.e. converges faster to higher fidelities) the basic Oxford one for all initial errors, while our superposed version clearly improves the double selection protocol for one of the errors ($\sigma_y$) and slightly outperforms it for initial Werner states, while remain the same (in terms of fidelity evolution) for the other two Pauli errors  ($\sigma_x, \sigma_z$), although the density matrix evolution differ.
\begin{figure}[ht]
    \centering
    \includegraphics[width=0.84\linewidth]{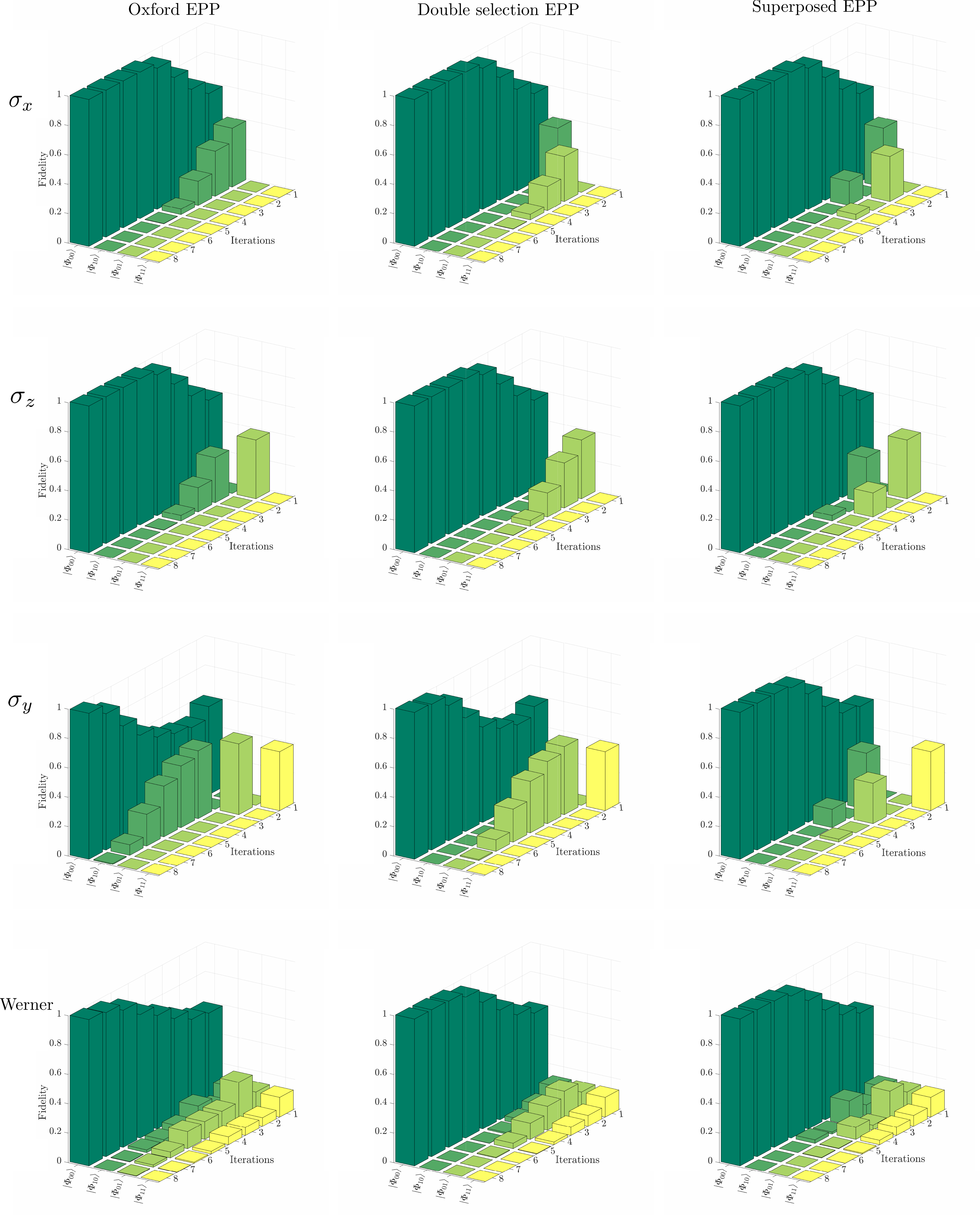}
    \caption{Pauli diagonal elements evolution for the different EPPs  for different initial Pauli errors and Werner states. }
    \label{fig:appendix}
\end{figure}

\end{document}